\documentclass[%
reprint,
showpacs,preprintnumbers,
nofootinbib,
amsmath,amssymb,
aps,
prd,
floatfix,
]{revtex4-1}

\usepackage{graphicx}
\usepackage{dcolumn}
\usepackage{bm}

\usepackage{url}

\newcommand{\boss}[2]{\ensuremath{\rlap{\kern-2.5pt\ensuremath{\overset{\scriptscriptstyle(-)}{\phantom{#1}}}}{\ensuremath{{#1}_{#2}}}}}

\begin{document}

\author{Carlo Giunti}
\email{giunti@to.infn.it}
\altaffiliation[also at ]{Department of Theoretical Physics, University of Torino, Italy}
\affiliation{INFN, Sezione di Torino, Via P. Giuria 1, I--10125 Torino, Italy}

\author{Marco Laveder}
\email{laveder@pd.infn.it}
\affiliation{Dipartimento di Fisica ``G. Galilei'', Universit\`a di Padova,
and
INFN, Sezione di Padova,
Via F. Marzolo 8, I--35131 Padova, Italy}

\date{\today}



\title{Hint of CPT Violation in Short-Baseline Electron Neutrino Disappearance}

\begin{abstract}
We analyzed
the electron neutrino data of the Gallium radioactive source experiments
and
the electron antineutrino data of the reactor Bugey and Chooz experiments
in terms of neutrino oscillations
allowing for a CPT-violating difference of the squared-masses and mixings of
neutrinos and antineutrinos.
We found that
the discrepancy between the disappearance of electron neutrinos
indicated by the data of the Gallium radioactive source experiments
and the limits on the disappearance of electron antineutrinos
given by the data of reactor experiments
reveal a positive
CPT-violating asymmetry of the effective neutrino and antineutrino mixing angles
(with a statistical significance of about $3.5\sigma$),
whereas the squared-mass asymmetry is practically not bounded.
\end{abstract}

\maketitle

The radioactive source experiments performed
by the
GALLEX
\cite{Anselmann:1995ar,Hampel:1998fc,1001.2731}
and
SAGE
\cite{Abdurashitov:1996dp,hep-ph/9803418,nucl-ex/0512041,0901.2200}
collaboration for testing the respective
Gallium solar neutrino detectors
revealed a disappearance of electron neutrinos
with energy $E$ of the order of 1 MeV at a distance $L$ of the order of 1 m.
Since the ratio $L/E$ is of the order of $10 \, \text{eV}^{-2}$,
the disappearance could be due to short-baseline oscillations
(see Ref.~\cite{Giunti-Kim-2007})
generated by a
squared-mass difference $\Delta{m}^2 \gtrsim 0.1 \text{eV}^2$
\cite{hep-ph/9411414,Laveder:2007zz,hep-ph/0610352,0707.4593,0711.4222,0902.1992,1005.4599,1006.2103,1006.3244}
and a large effective mixing angle $\vartheta$,
such that $\sin^22\vartheta \gtrsim 0.1$ \cite{1006.3244}.
On the other hand,
the measurements of
reactor electron antineutrino experiments constrain $\sin^22\vartheta$
below about 0.1
\cite{Declais:1995su,hep-ex/0301017,0711.4222},
assuming that the survival probabilities of neutrinos and antineutrinos are equal,
as implied by the CPT symmetry
(see Ref.~\cite{Giunti-Kim-2007}).

We can test the compatibility of electron neutrino
disappearance in Gallium radioactive source experiments
with the reactor constraints through a calculation of the corresponding
parameter goodness-of-fit \cite{hep-ph/0304176}.
We use the fit of Gallium data presented in Ref.~\cite{1006.3244}
and the fit of the Bugey \cite{Declais:1995su} and Chooz \cite{hep-ex/0301017}
reactor data presented in Ref.~\cite{0711.4222},
taking into account also the constraints given by the results of the
Mainz \cite{hep-ex/0412056}
and
Troitsk \cite{Lobashev:2003kt}
Tritium $\beta$-decay experiments
as described in Ref.~\cite{1005.4599}.
For the parameter goodness-of-fit (GoF) we obtain
\begin{equation}
\Delta\chi^2_{\text{min}}
=
12.1
\,,
\quad
\text{NDF}
=
2
\,,
\quad
\text{GoF}
=
0.2\%
\,,
\label{001}
\end{equation}
where
$\text{NDF}$ is the number of degrees of freedom
and
$\Delta\chi^2_{\text{min}}$
is the difference between the $\chi^2_{\text{min}}$
obtained in the combined analysis
and the sum of the $\chi^2_{\text{min}}$'s
obtained in the separate analyses of
Gallium data
and
reactor plus Tritium data.

Therefore,
electron neutrino
disappearance in Gallium radioactive source experiments
is rather incompatible with the reactor constraints
on the disappearance of electron antineutrinos
and we are lead to study the possibility of
CPT violation which can generate a difference of the survival probabilities of
neutrinos and antineutrinos.

CPT symmetry is widely believed to be exact,
because it is a fundamental symmetry of local relativistic Quantum Field Theory
(see Ref.~\cite{hep-ph/0309309}).
However,
it is possible to extend the Standard Model Lagrangian by including
CPT and Lorentz violating terms
\cite{hep-ph/9703464,hep-th/0012060,0801.0287}.

We are stimulated in considering CPT violation
by the recent indication in favor of CPT violation found
in the MINOS long-baseline $\nu_{\mu}$ and $\bar\nu_{\mu}$
disappearance experiment \cite{0910.3439,MINOS-Neutrino2010}.
The MINOS data indicate for $\nu_{\mu}$ and $\bar\nu_{\mu}$
different values of the effective squared-mass differences
and mixings.
Also the difference between the absence of
$\nu_{\mu}\to\nu_{e}$ oscillations in the data of the
short-baseline MiniBooNE experiment
\cite{0812.2243}
and the indication in favor
short-baseline $\bar\nu_{\mu}\to\bar\nu_{e}$ oscillations
which are compatible with the LSND signal
\cite{hep-ex/0104049}
found recently in the MiniBooNE experiment
\cite{MiniBooNE-Neutrino2010,1007.1150}
may be due to different values of the effective squared-mass differences
and mixings of neutrinos and antineutrinos \cite{hep-ph/0308299}.
Such difference in the fundamental properties of neutrinos and antineutrinos
are possible if the theory is nonlocal
\cite{hep-ph/0201258}.

Hence, we consider the simplest case in which
short-baseline disappearance of electron neutrinos and antineutrinos
are given by effective two-neutrino like oscillation probabilities
governed by different effective squared-mass differences
and mixings
\cite{hep-ph/0010178,hep-ph/0108199,hep-ph/0112226,hep-ph/0201080,hep-ph/0201134,hep-ph/0201211,hep-ph/0307127,hep-ph/0308299,hep-ph/0505133,hep-ph/0306226,0804.2820,0903.4318,0907.5487,0908.2993,1005.4599},
$\Delta{m}^{2}_{\nu}$ and $\sin^22\vartheta_{\nu}$ for neutrinos
and
$\Delta{m}^{2}_{\bar\nu}$ and $\sin^22\vartheta_{\bar\nu}$ for antineutrinos:
\begin{align}
P_{\nu_{e}\to\nu_{e}}
=
\null & \null
1 - \sin^22\vartheta_{\nu} \sin^2 \left( \frac{\Delta{m}^{2}_{\nu} L}{4 E} \right)
\,,
\label{002}
\\
P_{\bar\nu_{e}\to\bar\nu_{e}}
=
\null & \null
1 - \sin^22\vartheta_{\bar\nu} \sin^2 \left( \frac{\Delta{m}^{2}_{\bar\nu} L}{4 E} \right)
\,.
\label{003}
\end{align}
These survival probabilities can be obtained in a CPT-violating version of four-neutrino mixing schemes
(see Refs.~\cite{hep-ph/9812360,GonzalezGarcia:2007ib})
as hypothesized in Ref.~\cite{hep-ph/0308299}.
Four-neutrino mixing schemes
are the simplest extensions of the standard three-neutrino mixing schemes which can accommodate
the two measured small solar and atmospheric squared-mass differences
$
\Delta{m}^2_{\text{SOL}}
\simeq
8 \times 10^{-5} \, \text{eV}^2
$
and
$
\Delta{m}^2_{\text{ATM}}
\simeq
2 \times 10^{-3} \, \text{eV}^2
$
and one larger squared-mass difference for short-baseline neutrino oscillations,
$
\Delta{m}^2 \gtrsim 0.1 \, \text{eV}^2
$.
The existence of a fourth massive neutrino corresponds,
in the flavor basis,
to the existence of a sterile neutrino $\nu_{s}$.

\begin{figure}[t!]
\begin{center}
\includegraphics*[bb=22 147 564 698, width=\linewidth]{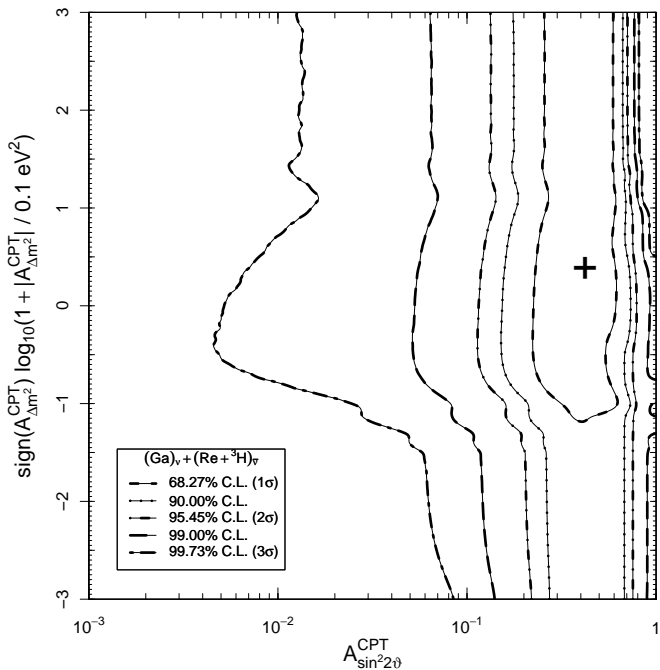}
\end{center}
\caption{ \label{006b}
Allowed regions in the
$A_{\sin^22\vartheta}^{\text{CPT}}$--$A_{\Delta{m}^{2}}^{\text{CPT}}$ plane.
The best-fit point corresponding to $\chi^2_{\text{min}}$ is indicated by a cross.
}
\end{figure}

We analyze
Gallium data
and
reactor plus Tritium data
in terms of the mass and mixing asymmetries
\begin{align}
A_{\Delta{m}^{2}}^{\text{CPT}}
=
\null & \null
\Delta{m}^{2}_{\nu} - \Delta{m}^{2}_{\bar\nu}
\,,
\label{004}
\\
A_{\sin^22\vartheta}^{\text{CPT}}
=
\null & \null
\sin^22\vartheta_{\nu} - \sin^22\vartheta_{\bar\nu}
\,.
\label{005}
\end{align}
The best-fit values of the asymmetries are
\begin{equation}
(A_{\sin^22\vartheta}^{\text{CPT}})_{\text{bf}}
=
0.42
\,,
\quad
(A_{\Delta{m}^{2}}^{\text{CPT}})_{\text{bf}}
=
0.37 \, \text{eV}^2
\,.
\label{007}
\end{equation}
The allowed regions at
68.27\%,
90\%,
95.45\%,
99\% and
99.73\%
C.L.
for
$A_{\sin^22\vartheta}^{\text{CPT}}$
and
$A_{\Delta{m}^{2}}^{\text{CPT}}$
are shown in Fig.~\ref{006b}.
We used a logarithmic scale for $A_{\sin^22\vartheta}^{\text{CPT}}$,
considering only the interval
$10^{-3} \leq A_{\sin^22\vartheta}^{\text{CPT}} \leq 1$
which contains all the allowed regions.
For $A_{\Delta{m}^{2}}^{\text{CPT}}$
we used an antisymmetric logarithmic scale,
which allows us to show both positive and negative values of
$A_{\Delta{m}^{2}}^{\text{CPT}}$,
enlarging the region of small values of $A_{\Delta{m}^{2}}^{\text{CPT}}$
between 0.1 and 1 eV$^2$.

The best-fit value
$(A_{\Delta{m}^{2}}^{\text{CPT}})_{\text{bf}}$
of the mass asymmetry
is small,
but Fig.~\ref{006b} shows that
in practice any value of the mass asymmetry is allowed,
with a slight preference for positive values.
On the other hand,
we obtain a very interesting result for the mixing asymmetry:
the best-fit value
$(A_{\sin^22\vartheta}^{\text{CPT}})_{\text{bf}}$
is large and positive and
Fig.~\ref{006b} shows that zero or negative values are disfavored.

\begin{figure}[t!]
\begin{center}
\includegraphics*[bb=24 147 564 702, width=\linewidth]{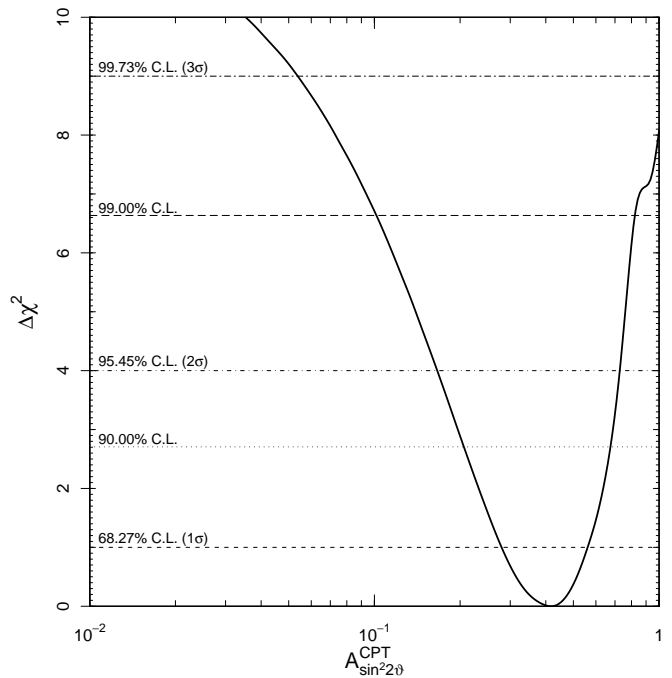}
\end{center}
\caption{ \label{008}
Marginal $\Delta\chi^{2}=\chi^{2}-\chi^{2}_{\text{min}}$
for
$A_{\sin^22\vartheta}^{\text{CPT}}$.
The $\Delta\chi^{2}$ of the horizontal lines correspond to the indicated value of confidence level.
}
\end{figure}

From Fig.~\ref{006b} one can see that
the smallest value of
$A_{\sin^22\vartheta}^{\text{CPT}}$
included in the $3\sigma$ allowed region
is about
0.005
at
$A_{\Delta{m}^{2}}^{\text{CPT}} \simeq -0.15 \, \text{eV}^2$.
However,
since in practice
$A_{\Delta{m}^{2}}^{\text{CPT}}$
is not bounded,
the statistically reliable limits on $A_{\sin^22\vartheta}^{\text{CPT}}$
are given by the marginal $\Delta\chi^{2}=\chi^{2}-\chi^{2}_{\text{min}}$
function for
$A_{\sin^22\vartheta}^{\text{CPT}}$
depicted in Fig.~\ref{008}.
One can see that
$A_{\sin^22\vartheta}^{\text{CPT}}>0.055$
at $3\sigma$.

The marginal $\Delta\chi^{2}$ of a null asymmetry
($A_{\sin^22\vartheta}^{\text{CPT}}=0$)
is
$12.0$,
with an associated p-value of
$0.05\%$.
Hence,
there is an indication of a positive asymmetry
$A_{\sin^22\vartheta}^{\text{CPT}}$
at a level of about $3.5\sigma$.

The indication in favor of a CPT asymmetry that we have found is robust,
because it is obtained by confronting the observations on the disappearance of electron
neutrino and antineutrino,
which should be equal if the CPT symmetry is not violated.
We considered the simplest case of a difference of
the effective squared-masses and mixings of
neutrinos and antineutrinos.
The analysis of the data in the framework of other, more complicated, models
would lead to a similar indication of a CPT asymmetry
in the space of the parameters of the specific model under consideration.

Our results depend on the hypothesis that
the anomalous deficit of
electron neutrinos measured in the radioactive source experiments
is due to neutrino oscillations
\cite{Laveder:2007zz,hep-ph/0610352,0707.4593,0711.4222,0902.1992,1005.4599,1006.2103,1006.3244},
taking into account the uncertainty of the detection cross section
\cite{nucl-th/9503017,hep-ph/9710491,nucl-th/9804011}
as discussed in Ref.~\cite{1006.3244}.
The experimental significance of the anomaly can be tested by the new
Gallium radioactive source experiment
proposed in Ref.~\cite{1006.2103}.
However,
a crucial improvement needed for understanding the validity of
the neutrino oscillation hypothesis is an accurate calculation of the
$\nu_{e}$-${}^{71}\text{Ga}$ detection cross section
and its uncertainty,
improved with respect to the existing ones
\cite{hep-ph/9710491,nucl-th/9804011}.

The short-baseline disappearance of electron neutrinos
can be tested in the future not only with new
Gallium radioactive source experiments,
but also with accelerator experiments with a well-known flux of electron neutrinos,
as discussed in Ref.~\cite{1005.4599}.

For the investigation of the CPT asymmetry,
the ideal experiments are those which can measure the disappearance of both
electron neutrinos and antineutrinos,
with sources which emit well-known neutrino and antineutrino fluxes
and detection processes with well-known cross sections.
Experiments of this type are near-detector
beta-beam \cite{0907.3145}
and
neutrino factory \cite{0907.5487,1005.3146}
experiments,
which are under study but may require a long time to be realized.
In a shorter time it may be possible to perform dedicated experiments with
intense artificial radioactive sources of electron neutrinos and antineutrinos
placed inside or close to neutrino elastic scattering detectors with a low energy threshold,
as Borexino \cite{Bellotti-private-10} or a low-threshold liquid Argon TPC
\cite{hep-ex/0103008}.

In conclusion,
we have found an indication of a CPT-violating asymmetry in the short-baseline disappearance of electron neutrinos and antineutrinos
by confronting the neutrino data of the Gallium radioactive source experiments
and the antineutrino data of the reactor Bugey and Chooz experiments.
Considering the simplest case of a difference of squared-masses and mixings of
neutrinos and antineutrinos,
we found that the squared-mass asymmetry is practically not bounded,
whereas the mixing asymmetry is positive
with a statistical significance of about $3.5\sigma$.

%

\end{document}